\documentclass[%
 reprint,
 amsmath,amssymb,
 aps,
]{revtex4-2}
\usepackage{graphicx}
\usepackage{dcolumn}
\usepackage{bm}
\usepackage{xcolor} 
\usepackage{hyperref}

\newcommand\aap{Astron.~Astrophys.}
\newcommand\apjl{Astrophys.~J.~Lett.}
\newcommand\apjs{Astrophys.~J.~Suppl.~Ser.}
\newcommand\araa{Annu.~Rev.~Astron.~Astrophys.}
\newcommand\mnras{Mon.~Not.~R.~Astron.~Sco.}
\newcommand\jcap{J.~Cosmo.~Astropart.~Phys.}
\newcommand\sovast{Sov.~Astron.}

\begin{document}

\preprint{APS/LV16677}

\title{Bondi-Hoyle-Lyttleton Accretion in a Reactive Medium: Detonation Ignition \\ and a Mechanism for Type Ia Supernovae}

\author{Heinrich Steigerwald}
\email{heinrich.steigerwald@ufes.br}
\affiliation{Center for Astrophysics and Cosmology (Cosmo-ufes) and Department of Physics, Federal University of Esp\'{i}rito Santo, 29075910 Vit\'{o}ria, Esp\'{i}rito Santo, Brazil.}
\author{Emilio Tejeda}%
\affiliation{C\'atedras CONACyT--Instituto de F\'isica y Matem\'aticas, Universidad Michoacana 
de San Nicol\'as de Hidalgo, Edificio C-3, Ciudad Universitaria, 58040 
Morelia, Michoac\'an, Mexico}

\date{November 11, 2020}

\begin{abstract}
Detonation initiation in a reactive medium can be achieved by an externally created shock wave. Supersonic flow onto a gravitating center, known as Bondi-Hoyle-Lyttleton (BHL) accretion, is a natural shock wave creating process, but, to our knowledge, a reactive medium has never been considered in the literature.
Here, we conduct an order of magnitude analysis to investigate under which conditions the shock-induced reaction zone recouples to the shock front. We derive three semianalytical criteria for self-sustained detonation ignition.
We apply these criteria to the special situation where a primordial black hole (PBH) of asteroid mass traverses a carbon-oxygen white dwarf (WD). 
Since detonations in carbon-oxygen WDs are supposed to produce normal thermonuclear supernovae (SNe Ia), the observed SN Ia rate constrains the fraction of dark matter (DM) in the form of PBHs as 
$\log_{10}(f_{\rm PBH})< 0.8 \log_{10}(M_{\rm BH}/3\times 10^{22}{\rm g})$
in the range $10^{21}\!-\!10^{22}$g ($10^{20}\!-\!10^{22}$g) from a conservative (optimistic) analysis. 
Most importantly, these encounters can account for both the rate and the median explosion mass of normal sub-Chandrasekhar SNe Ia if a significant fraction of DM is in the form of PBHs with mass $10^{23}$g. 
\end{abstract}

\maketitle


\section{Introduction}
Detonation ignition in a reactive medium can be achieved by a sufficiently strong shock wave, the requirement, known as Zeldovich's criterion \citep{zeldovich1956soviet}, being that the shock velocity must surpass the Chapman-Jouguet (CJ) velocity over a distance of at least the induction length (see also Ref.~ \citep{body1997mechanism}).

For a wedge-induced shock wave, an oblique detonation can be ignited if the preshock flow velocity exceeds the CJ velocity and the orthogonal postshock velocity is at most sonic  \citep{gross1963}. Then, a structure composed of four elements is observed: a nonreactive oblique shock, an induction zone, a set of deflagration waves, and a reactive shock where the front is intimately linked to the energy release  \citep{1994Li,VIGUIER19963023}. For high wedge angles, where part of the postshock flow is subsonic, the detonation detaches and travels upstream  \citep{1994Li}.

Bondi-Hoyle-Lyttleton (BHL) accretion, the process where a gravitating center with mass $M$ is placed in a continuous supersonic flow, produces a bow shock as a result of gravitational focusing of streamlines in the wake of the accretor.
The relevant length scale is the critical impact parameter $b_c$, inside of which all matter is accreted onto the gravitating center, 
\begin{align}\label{eq:bc}
b_c \simeq  \frac{2\;\! G \:\! M}{v_{\infty}^{\,2}}\,,
\end{align}
where $v_{\infty}$ is the relative velocity far from the accretor, and $G$ the gravitational constant. To our knowledge, accretion from a reactive medium has never been considered in the literature. One objective of this work is to fill this gap.

The second objective, as direct application of the first, is a critical reanalysis of the situation, considered initially by  \citet{2015PhRvD..92f3007G} (hereafter G15), where the passage of a primordial black hole (PBH) through a carbon-oxygen white dwarf (WD) initiates thermonuclear runaway through localized dynamical friction heating. This evokes the possibility of triggering thermonuclear supernovae (SNe Ia).

PBHs in the asteroid mass window ($10^{20}\!-\!10^{23}$g) are compelling candidates for dark matter (DM), 
since they are formed naturally at the end of inflation 
\citep{2016PhRvD..94f3509K,2016PhRvD..94h3523K,2017PhRvD..95l3510I,2019JCAP...01..037D,2020PhRvD.101b3513L,2020PDU....2700440G,2020EPJC...80..917A,2021JCAP...01..040O} and still resist all observational constraints  \cite{2019JCAP...08..031M,2020PhRvD.101f3005S,2020ARNPS..70..355C,2021JPhG...48d3001G}. Given their potentially huge number density, encounters with stellar objects should be frequent enough to be relevant on cosmological timescales. 

Thermonuclear supernovae, though commonly thought to explode in binary systems, are consistent with detonations of single WDs with median mass $1 M_{\odot}$   \cite{2010ApJ...714L..52S}. Intriguingly, a high percentage of massive single WDs originates from main-sequence or post-main-sequence mergers  \citep{2018MNRAS.479L.113K}. 
These have carbon-oxygen cores and are potential progenitors for PBH triggers of SNe Ia.

A careful analysis of the work of G15 and the follow-up study of  \citet{2019JCAP...08..031M} (hereafter M19), shows that the investigated criteria match with triggers of deflagrations (subsonic combustion waves). Given the low single WD core temperature ($10^7\,$K) and the sharp temperature profile created by the passage of the PBH, these deflagrations cannot transition to detonation, as is possible in near-Chandrasekhar preheated WDs \citep{2009ApJ...696..515S}. They would, therefore, eject large portions of unburnt carbon and oxygen, which is unobserved in normal SN Ia spectra \citep{2006ApJ...645.1392M}. 
The second objective of the present Letter is to consider direct detonation ignition from BHL accretion.

The Letter is organized as follows: In Sec.~\ref{sec:RBHLpre}, we briefly review the BHL preshock flow, in Sec.~\ref{sec:RBHLshock} we analyze the shock geometry and in Sec.~\ref{sec:RBHLcrit} we derive general criteria for detonation ignition. In Sec.~\ref{sec:SN}, we reinvestigate PBH triggers of SNe Ia, presenting in Sec.~\ref{sec:SNmod} our WD modeling, in Sec.~\ref{sec:SNcross} the ignition cross-section and in Sec.~\ref{sec:SNrate} the resulting SN Ia event rate and explosion mass. In Sec.~\ref{sec:conclusion}, we present our conclusions.

\section{Reactive Bondi-Hoyle-Lyttleton accretion}\label{sec:RBHL}

\subsection{Pre-shock flow equations}\label{sec:RBHLpre}
\begin{figure}[htb]
\includegraphics[width=\linewidth]{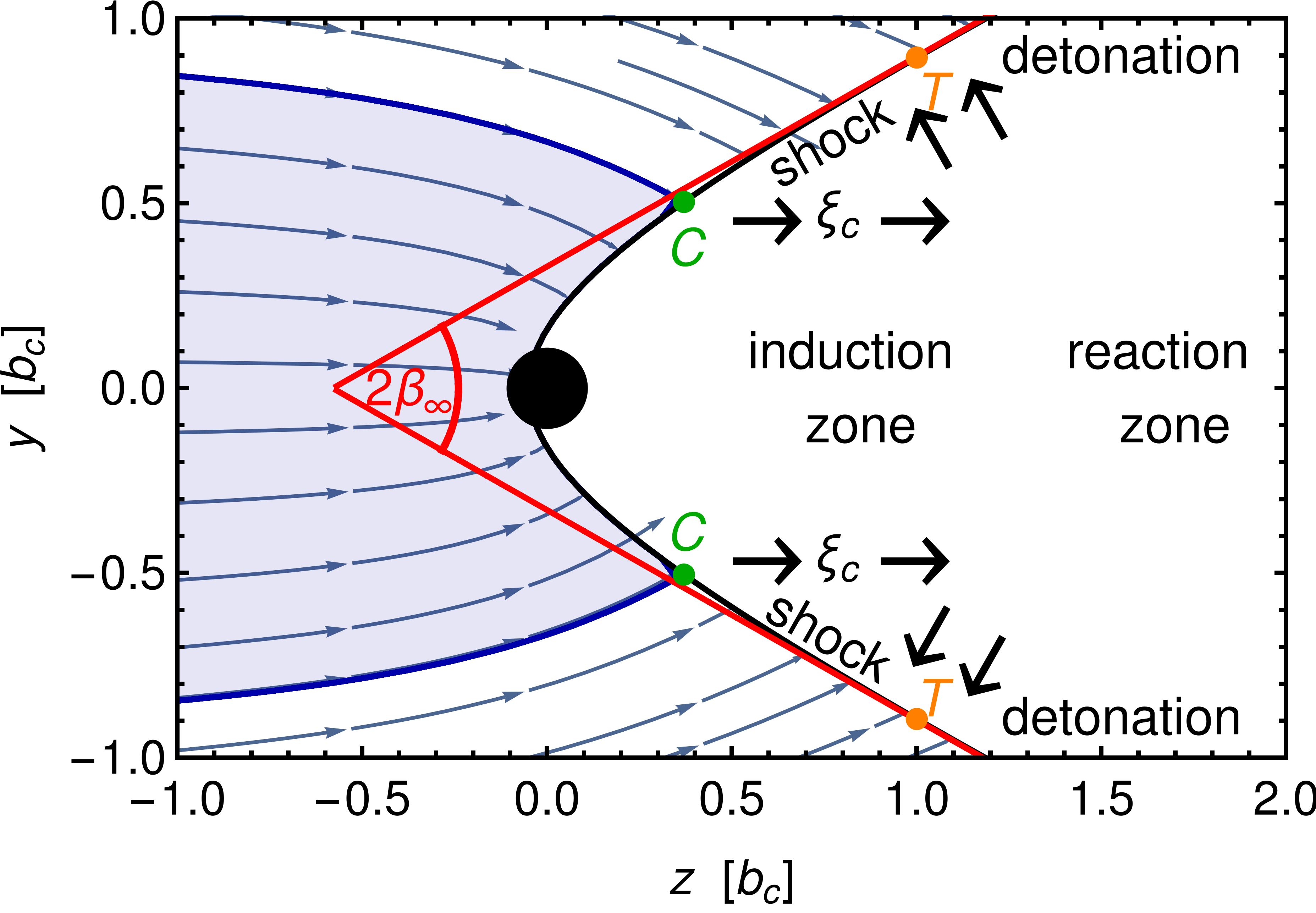}
\caption{Schematic profile view of reactive BHL flow, entering from the left $(\theta=\pi)$. Matter inside the shaded area is ultimately accreted onto the central object (black disk, largely enhanced for illustrative purposes). Postshock flow is omitted for clarity. Letter $C$ ($T$) denotes the critical (triple) point.
}
\label{fig:1}
\end{figure}

Consider spherical coordinates ($r$, $\theta$, $\phi$) with the accretor at rest at the origin. A homogeneous ideal fluid is entering from the south pole ($\theta=\pi$) along the $z$ axis with velocity $v_{\infty}$ (see Fig.~\ref{fig:1}). Axial symmetry implies that the equations are independent of $\phi$. The fluid has density $\rho$, pressure $p$, specific heat ratio $\gamma$, sound speed $a=\sqrt{\partial p/\partial\rho}$, and Mach number $\mathcal{M} = v/a$. 
Indices $\infty$, $1$, and $2$ denote ``far from the accretor", ``immediateley before the shock", and ``immediately after the shock", respectively. For highly supersonic inflow ($\mathcal{M}_{\infty}^{\,2} \gg 1$), the stream field around the accretor is well described in ballistic approximation (see, for example, Refs.~\citep{1957MNRAS.117...50D,1979SvA....23..201B}):
\begin{align}
v_r =&\;  v_{\infty} \left(\!1+\frac{b_c}{r}- \frac{b^2}{r^2}\right)^{\!\!1/2}\!\!\!\! {\rm sgn}(\theta_p\!-\theta)\,, \label{eq:vr}\\
v_{\theta}= &\; -v_{\infty}\:\! \frac{b}{r}\,, \label{eq:vtheta}\\
\rho =&\; \frac{\rho_{\infty}\;\!b^2}{r\sin\theta\;\!(2\;\!b-r\sin\theta)}\,,\label{eq:rho}\\
r=&\; \frac{2\;\!b^2}{b_c(1+\cos\theta)+2\;\!b\sin\theta}\,,  \label{eq:r}\\
\theta_p =&\; \arccos\left[\left(1+\frac{4\:\!b^2}{b_c^{\,2}}\right)^{\!\!-1/2}\right]\,, \label{eq:thetap}
\end{align}
where $v_r$ and $v_{\theta}$ are radial and polar velocity components, respectively, of a given fluid particle with impact parameter $b$ and related to the total velocity by $v^2 = v_r^{\,2}+v_{\theta}^{\,2}$, $b_c$ is given by Eq.~\eqref{eq:bc}, $\theta_p$ is the periastron angle, and  sgn$(x)$ the sign function \footnote{${\rm sgn}(x)=-1$ if $x<0$, sgn$(x)=0$ if $x=0$ and sgn$(x)=1$ if $x>0$}. Streamlines of this flow are shown in Fig.~\ref{fig:1}. %

\subsection{Nonreactive shock geometry}\label{sec:RBHLshock}
The BHL shock is different from a classical bow shock produced by a blunt body in that both preshock Mach number $\mathcal{M}_{1}$ and density $\rho_1$ vary along the front. 
The incoming orthogonal flow velocity at any point ($r,\theta$) on the shock is given by
\begin{equation}\label{eq:vperp1}
  v_{\!\perp 1} = -\cos (\theta\!-\!\beta) \, v_{\theta} - \sin(\theta\!-\!\beta)\, v_{r},
\end{equation}
where $\beta$ is the shock opening half-angle (the angle between the tangent to the shock and the $z$ axis) and $v_{r}$ and $v_{\theta}$ are specified by Eqs.~(\ref{eq:vr}) and (\ref{eq:vtheta}), respectively. 

If the shock geometry is a known parametric function of the form $y(z)$, then the opening half-angle is
\begin{equation}
    \tan\beta = \frac{dy}{dz}.
\end{equation}
In general, $y(z)$ depends on $\mathcal{M}_{\infty}$, $\gamma$, and the size of the accretor $r_{*}$
 \cite{1997A&A...320..342F}. 
By performing numerical simulations of wind accretion flow onto a point mass,  we find that for a polytropic fluid around a small accretor ($r_{*} \ll b_c$), the bow shock is well approximated by the hyperbola,
\begin{equation}
    y(z) \simeq \tan\beta_{\infty} \sqrt{(z_{\infty}\!+z)^2-(z_{\infty}\!-\!z_{0})^2},
    \label{eq:fit}
\end{equation}
where $\beta_{\infty}$ is the asymptotic half-opening angle, $-z_{\infty}$ is the apex of the asymptotic shock cone, and $-z_0$ the offset of the shock at $y=0$ (see Fig.~\ref{fig:1}). In 
App.~\ref{sec:aztekas}
we provide fitting formulas for the coefficients 
$\beta_{\infty}$, $z_{\infty}$ and $z_{0}$
for the parameter ranges $1.3 \leq \gamma \leq 1.7$ and $2.5\leq \mathcal{M}_{\infty}\leq 8$.

\subsection{Detonation formation}\label{sec:RBHLcrit}
The propagation velocity of detonation $v_{\rm CJ}$ is 
entirely determined by the reactions' total specific heat release $q$. For a perfect fluid and high Mach numbers, $v_{\rm CJ} \simeq \sqrt{2\;\! (\gamma+\!1)\;\! q}$ (see, for example, Ref.~ \citep{ciccarelli2008flame}). 
According to the Zeldovich-von Neuman-Doring (ZND) model, the wave structure is a lead shock followed by a reaction zone. These are separated by the induction length $\xi$. 
In a self-sustained detonation, pressure waves (shown as double arrows in Fig.~\ref{fig:1}) created by the heat release in the reaction zone can reach the shock and power it.

Since $v_{\perp 1}$ and $\rho_{1}$ are decreasing functions of $r$ and $b$,
reactions are more propitious to occur close to the "critical point" $C$ (see Fig.~\ref{fig:1}), remembering that the gas crossing the shock inside $C$ accretes onto the central object. Using Eq.~\eqref{eq:r} and the shock function $y(z)$, the coordinates $(z_c,y_c)$ of the critical point are solution to the equation
\begin{equation}
    b_c \Big[\!\sqrt{y(z)^2+z^2}+z\Big] +2\;\!b\;\!y(z)-2\;\!b^2 = 0
\end{equation}
with $b=b_c$.

Let us note $\xi_c$ the ``critical induction length", calculated for the preshock conditions at $C$. An inspection of the nonreactive downstream flow at $C$ reveals that streamlines are approximately parallel to the $z$ axis (see
Fig.~\ref{fig:S1}).
Therefore, the reaction zone sets in at an effective distance $\xi_c=\xi_{\perp c}/\sin\beta_c$ up the $z$ axis, where $\xi_{\perp c}$ is computed with the one-dimensional ZND model 
(see Ref.\cite{1999ApJ...512..827G} and App.~\ref{sec:ZND}),
and the preshock conditions at $C$.
The orthogonal projection from the reaction zone on the shock front defines the ``triple point"  $T$ (see Fig.~\ref{fig:1}), with coordinates $(z_t,y_t)$ given by
\begin{equation}
\frac{dy}{dz}(z_t,y_t) = \frac{z_c+ \xi_c-z_t}{y_t-y_c}.    
\end{equation}

At this point, the oblique shock can transition to an oblique detonation if the reaction zone is sonically connected with the shock front. 

First, in virtue of the oblique detonation ignition criteria \cite{gross1963}, the postshock orthogonal Mach number at the end of the critical induction length must be at most sonic:
\begin{align}\label{eq:crit1}
    \mathcal{M}_{\!\perp 2}(r_t,\theta_t) < 1.
\end{align}

Second, since the total preshock veloctiy $v_1$ is a decreasing function of $r$, if it exceeds the CJ velocity at the triple point, then it also does so over the whole induction length. Therefore, we must have
\begin{align}\label{eq:crit2}
    v_{1}(r_t,\theta_t) > v_{\rm CJ}.
\end{align}

Finally, the postshock state is spatially and temporally limited as it fades out at large radii as a Mach wave, and rarefies after the passage of the shock through an expansion wave.  To ensure that reactions occur during the high-pressure post-shock state, 
the critical induction length must be at most of the order of the critical impact parameter,
\begin{align}\label{eq:crit3}
    \xi_{\perp c} < \alpha \;\! b_c,
\end{align}
where $\alpha$ is a parameter of order unity. Since determining $\alpha$ exceeds the scope of the present work, in the next section we will consider two cases: $\alpha=1$ (conservative) and $\alpha=10$ (optimistic).
Equations~\eqref{eq:crit1}--\eqref{eq:crit3} constitute the three semianalytical criteria necessary for successful shock-induced, self-sustained detonation ignition in a BHL flow.

\section{PBH triggers of SNe Ia}\label{sec:SN}
In this section, we investigate if PBHs can ignite detonations in WD cores according to criteria (\ref{eq:crit1})--(\ref{eq:crit3}), considering PBHs in the mass range $10^{19}\!-\!10^{25}$g and WDs in the mass range $0.6 M_{\odot}\! - \! 1.4 M_{\odot}$. 

\subsection{WD modeling}\label{sec:SNmod}
The velocity of a PBH relative to the WD matter is the sum of the mean galactic velocity and the free-fall velocity. Neglecting the galactic velocity, the relative velocity inside the WD at a radial position $R$ from the center of the WD is
\begin{align}\label{eq:v}
v_{\infty}(R) = c\;\! \Big[1-\exp\Big(\frac{2\;\! \Phi(R)}{c^2}\Big)\Big]^{1/2}\,,
\end{align}
where $\Phi(R)$ is the gravitational potential of the WD normalized by $\Phi(\infty)=0$ and $c$ is the light speed.

We model the initial WD state as a zero temperature ideal Fermi gas of degenerate electrons. 
The mass-radius relation is obtained from the numerical integration of the Tolman-Oppenheimer-Volkoff equation, and the gravitational potential is obtained from numerical integration of the Poisson equation.

For the considered parameter space, the asymptotic adiabatic index is $1.33\lesssim \gamma_{\infty} \lesssim 1.57$ and the asymptotic free-fall Mach number is $2.6\lesssim \mathcal{M}_{\infty}\lesssim 7.7$. 
We determine the relevant shock geometry according to Sec.~\ref{sec:RBHLshock} in the polytropic approximation. A more realistic shock modeling with the hot WD equation of state exceeds the scope of the present Letter and is the objective of an upcoming work.

The shocked state requires full nonzero temperature treatment of the equation of state. For the electron-positron part, we use the Nadyozhin equation of state (see Ref.~ \cite{1996ApJS..106..171B} for a very detailed review), which is well adapted for the double-transitional regime of a degenerate (nondegenerate) and relativistic (nonrelativistic) electron-positron gas \cite{1999ApJS..125..277T}.
Our code includes ideal gas ions and radiation and is tested for thermodynamical consistency (see Ref.~ \cite{1999ApJS..125..277T} for details).

In carbon-oxygen matter, the nuclear reactions can be categorized in three major exothermic stages: the fusions of carbon, oxygen, and silicon. These stages are spatially separated, such that the heat release of carbon fusion entirely governs the initial detonation propagation.

We determine the carbon induction length by integrating the Zeldovich-von Neuman-Doring equations adopting the specific nuclear energy generation rate $\dot{q}$ of  \cite{1988ADNDT..40..283C} including also electron screening (see, for example, Ref.~ \cite{2019JCAP...08..031M} for details). The initial conditions are given by the postshock state ($\rho_2,T_{2}$), which is determined by solving the shock jump conditions with the help of the Nadyozhin equation of state (see App.~\ref{sec:ind}
for details).
For corresponding parameters, our computed induction lengths agree with the results of Ref.~\cite{1999ApJ...512..827G}. We use the density dependent CJ velocity calculated by Ref.~\cite{1999ApJ...512..827G}.

\subsection{Ignition cross section}\label{sec:SNcross}

\begin{figure}[thb]
\includegraphics[width=\linewidth]{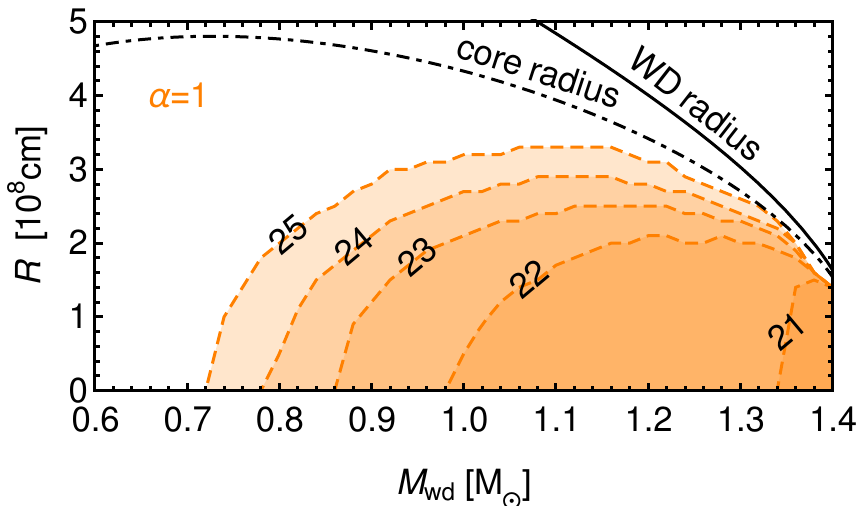}
\caption{Maximum radius for detonation ignition $R_m$ for $\log_{10}(M_{\rm BH}/{\rm g})$ as indicated by numbers and assuming $\alpha=1$. The core radius is defined by $\rho  \simeq 10^6\,$g$\,$cm$^{-3}$.
}
\label{fig:cross}
\end{figure}

Figure~\ref{fig:cross} shows the maximum radius $R_m$ where detonation ignition occurs according to criteria \eqref{eq:crit1}$\!-\!$\eqref{eq:crit3}. Criterion \eqref{eq:crit1} is always satisfied. For $M_{\rm BH}>10^{23}\,$g, criterion (\ref{eq:crit2}) is more stringent than (\ref{eq:crit3}), while for $M_{\rm BH}<10^{23}\,$g, it is the opposite. 
As expected, the ignition cross section $\pi R_m^2$ increases with PBH mass. Its dependence on WD mass is more subtle: since heavier WDs are smaller, the maximum is around $1.1 M_{\odot}$ for the heaviest PBHs considered, and increases to $1.4 M_{\odot}$ for the lightest. 
In App.~\ref{sec:fitting-cross},
we provide a fitting formula for the cross section.

\begin{figure}[htb]
\includegraphics[width=\linewidth]{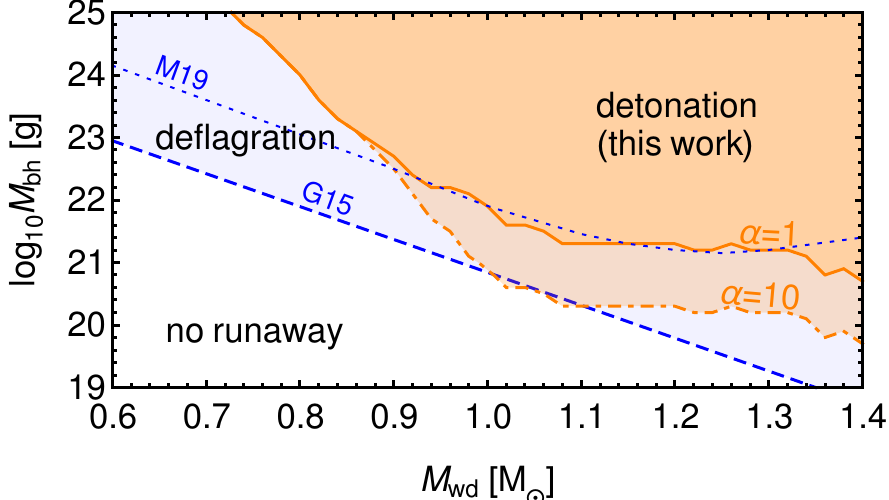}
\caption{Minimum PBH mass for direct detonation ignition as function of WD mass for passages through the center for $\alpha=1$ (orange full line) and $\alpha=10$ (orange dotdashed line), compared to minimum PBH mass for deflagration ignition according to G15 (blue dashed line) and M19 (blue dotted line).
}
\label{fig:comb}
\end{figure}

Figure~\ref{fig:comb} shows the minimum PBH mass for passages through the center. PBHs lighter than $10^{21}$g ($10^{20}$g) cannot ignite detonations if $\alpha=1$ ($\alpha=10$). 
Increasing (lowering) $\beta_{\infty}$ by $5^{\circ}$ lowers (increases) the minimum PBH mass by approximately $0.2\,$dex. Similarly, augmenting (decreasing) $v_{\rm CJ}$ by 10\% requires a $0.5\,$dex more (less) massive PBH. 

Figure~\ref{fig:comb} also compares our bounds with those of G15 and M19. 
As expected, our constraints are more stringent and complementary: 
given $M_{\rm BH}$ and $M_{\rm wd}$, three different outcomes are possible corresponding to the modes of combustion: detonation, deflagration, or unsuccessful runaway. For $\alpha=10$, slight overlapping occurs. The overlapping with the bounds of M19 can be explained by the circumstance that modeling of M19 excludes a cylindrical region of radius $b_c$, which is where, in our modeling, the induction zone in the wake of the accretor sets the conditions for detonation.

\subsection{SN Ia rate and median explosion mass}\label{sec:SNrate}

The encounter rate with detonation ignition is given by $[\pi \;\! R_m^{\,2}\;\! v_{\infty}(R_m)^2\;\! \rho_{\rm DM}\;\!f_{\rm PBH}]/[v_{\rm gal}\;\! M_{\rm PBH}]$, where $v_{\rm gal}$ is the mean galactic velocity, $\rho_{\rm DM}$ the dark matter density and $f_{\rm PBH}$ the fraction of dark matter in the form of PBHs. We consider $f_{\rm PBH}=1$ throughout.

Although a detonation can be triggered in the full range of WD masses considered, combustion of about $0.5 M_{\odot}$ of WD material into $^{56}$Ni, as indicated by basic SN Ia energetics and observations, requires a WD core density of $2\times 10^7\,$g$\,$cm$^{-3}$, which exists only in WDs of at least $0.85 M_{\odot}$ \cite{2010ApJ...714L..52S}.

To estimate the SN Ia rate, we adopt the local 100$\,$pc volume limited WD mass function of Ref.~\cite{2020ApJ...898...84K}, along with the estimate that our Galaxy contains $10^{10}\,$WDs  \cite{2009JPhCS.172a2004N}. 
We adopt the local DM density $\rho_{\rm DM}\approx 0.4\,$GeV$\,c^{-2}\,$cm$^{-3}$
and $v_{\rm gal} \approx 200\,$km$\,$s$^{-1}$.

\begin{figure}[htb]
\includegraphics[width=\linewidth]{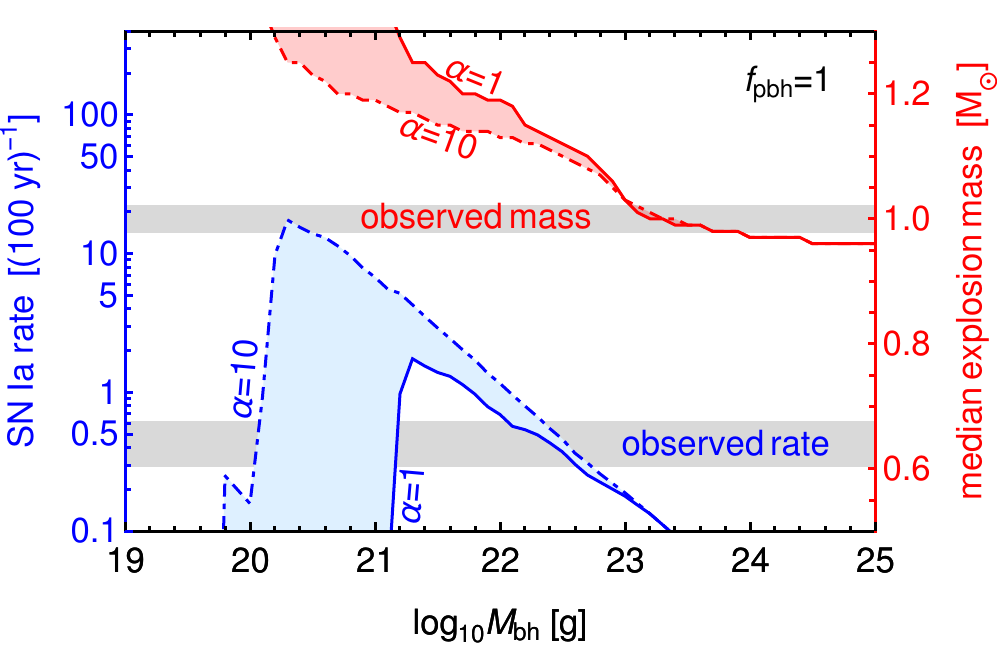}
\caption{SN Ia rate per century for the Milky Way (blue) and median explosion mass (red) as functions of monochromatic PBH mass, assuming all DM in the form of PBHs and adopting the local WD mass function. Further assumptions are discussed in the text. Note that the median explosion mass concerns ``normal" events while the shown rate concerns all SNe Ia. Considering that ``normal" events make up $\sim 70\%$ of all SNe Ia, PBHs with mass $\sim 10^{23}\,$g are roughly consistent with both the observed rate and median explosion mass. If only sub-Chandrasekhar events ($\sim 50\%$ of normal SNe Ia) are considered, the slight remaining tension is completely relieved.}
\label{fig:rate}
\end{figure}

Figure~\ref{fig:rate} shows the SN Ia rate as function of (monochromatic) PBH mass and for $f_{\rm PBH}=1$. 
Since the observed rate per century of $0.3\!-\!0.6$ \cite{1991ARA&A..29..363V,2011MNRAS.412.1473L} cannot be surpassed, the fraction of DM in the form of PBHs is constrained by 
$\log_{10}(f_{\rm PBH})< 0.8 \log_{10}(M_{\rm BH}/3\times 10^{22}{\rm g})$
in the range $10^{21}\!-\!10^{22}$g ($10^{20}\!-\!10^{22}$g) for $\alpha=1$ ($\alpha=10$). On the other hand, PBHs with either $10^{21}$g ($10^{20}$g) for $\alpha=1$ ($\alpha=10$) or, independently of $\alpha$, $10^{23}\,$g can account for 
normal SNe Ia, considering that normal events make up $70\%$ of the total rate. 

Figure~\ref{fig:rate} also shows the median explosion mass as function of $M_{\rm BH}$, adopting the same assumptions as for the SN Ia rate.
Since this should be around $1 M_{\odot}$  \citep{2010ApJ...714L..52S}, $10^{23}$g PBHs surprisingly match with both the rate and median explosion mass ($1.03 M_{\odot}$). 

On the other hand, PBHs with mass smaller than $10^{22}$g yield a median explosion mass larger than $1.2 M_{\odot}$ ($1.15 M_{\odot}$) for $\alpha=1$ ($\alpha=10$). This tightens further the aforementioned constraints on $f_{\rm PBH}$ and excludes these PBHs to account for sub-Chandrasekhar SNe Ia. 

There are important caveats to the rate and median explosion mass estimations. 
First, the measurement of the WD mass distribution is done in the solar neighborhood, while the estimate of the total number of WDs and type Ia SNe is done for the whole Galaxy. Second, the high mass tail of the WD mass distribution is uncertain. 
Third, about $10\%\!-\!30\%$ of all single WDs formed as a result of main-sequence or post-main-sequence mergers in binary systems  \cite{2017A&A...602A..16T}, consistent with binary population synthesis  \cite{2020A&A...636A..31T}. Since this percentage increases with increasing WD mass, we have assumed that all WD cores considered are carbon-oxygen. However, this assumption depends on the relatively unknown fraction of WDs heavier than $1.05 M_{\odot}$ that formed through single star evolution and are, therefore, oxygen-neon-magnesium. 
At the time of writing, it is suggested that CO WD merger remnants with mass $>1.06 M_{\odot}$ convert to ONe  \cite{2021ApJ...906...53S}. This would have implications for lower mass PBHs and merits to be investigated further.

\section{Conclusions}\label{sec:conclusion}
We have developed a semianalytical theory for reactive BHL flow.
The criteria for self-sustained detonation initiation are (1) the preshock flow velocity must exceed the CJ velocity at the triple point, (2) the postshock orthogonal Mach number at the triple point must be at most sonic, and (3) the critical induction length must be smaller than the critical impact parameter times a proportionality factor $\alpha$, where $\alpha=1$ ($\alpha=10$) in a conservative (optimistic) analysis.

In the second part, we have reanalyzed the G15 mechanism with the following conclusions. When an asteroid mass PBH passes through a carbon-oxygen WD, the BHL shock can lead to direct ignition of detonation (Fig.~\ref{fig:cross}). The parameter requirements are slightly more demanding than for deflagration ignition (see Fig.~\ref{fig:comb}). 
From the observed SN Ia rate, the fraction of DM in the form of PBHs is constrained by 
$\log_{10}(f_{\rm PBH})< 0.8 \log_{10}(M_{\rm BH}/3\times 10^{22}{\rm g})$
in the range $M_{\rm BH}>10^{21}$g ($M_{\rm BH}>10^{20}$g) for $\alpha=1$ ($\alpha=10$).
However, these constraints depend on the composition of WDs with  $M_{\mathrm{wd}}>1.06 M_{\odot}$ and more research is necessary.

Most importantly, in this work we have found that, independent of $\alpha$, and almost independent of the composition of WDs with $M_{\mathrm{wd}}>1.06 M_{\odot}$, PBHs with mass around $10^{23}$g can account for both the observed rate and median explosion mass of normal sub-Chandrasekhar SNe Ia (Fig.~\ref{fig:rate}). 
These PBHs could be detected or excluded in the very near future from the wave optics effect on microlensing \cite{2020MNRAS.493.3632S}.

\begin{acknowledgements}
The authors thank the anonymous referees for constructive comments that helped to improve this Letter. We thank Mukremin Kilic for kindly providing the local WD mass distribution. We thank Davi Rodriguez, Mukremin Kilic, Vadim Gamezo, Kevin Moore, and Alejandro Aguayo-Ortiz for useful discussions. H.S. is grateful for FAPES/CAPES DCR Grant No. 009/2014.
\end{acknowledgements}

\bibliography{biblio.bib}

\appendix

\section{BHL wind accretion simulations}\label{sec:aztekas}
To determine the non-reactive shock geometry, we perform simulations using the publicly available, numerical hydrodynamics code {\tt aztekas} with a polytropic ideal gas equation of state (see, for example, \cite{2019MNRAS.487.3607T,2019MNRAS.490.5078A}). 

\begin{table}[h]
\caption{\label{tab:1}%
Set of simulations and resulting parametrization of the steady-state, bow shock geometry.
}
\begin{ruledtabular}
\begin{tabular}{ccccc}
\textrm{$\mathcal{M}_{\infty}$} &
\textrm{$\gamma$} &
\textrm{$\tan(\beta_{\infty})$} &
\textrm{$z_{\infty}$} &
\textrm{$z_0$} \\ 
\colrule 
2.5 & 1.3 & 0.631 & 0.769 & -0.052 \\ 
2.5 & 1.4 & 0.654 & 0.976 & -0.006 \\ 
2.5 & 1.5 & 0.657 & 1.331 & 0.034 \\ 
2.5 & 1.6 & 0.664 & 1.638 & 0.083 \\ 
2.5 & 1.7 & 0.676 & 1.887 & 0.129 \\ 
5.0 & 1.3 & 0.319 & 3.015 & -0.105 \\ 
5.0 & 1.4 & 0.493 & 0.969 & -0.082 \\ 
5.0 & 1.5 & 0.416 & 2.156 & 0.023 \\ 
5.0 & 1.6 & 0.430 & 2.831 & 0.047 \\ 
5.0 & 1.7 & 0.450 & 3.015 & 0.088 \\ 
8.0 & 1.3 & 0.431 & 0.285 & -0.076 \\ 
8.0 & 1.4 & 0.410 & 0.977 & -0.053 \\ 
8.0 & 1.5 & 0.398 & 1.982 & 0.044 \\ 
8.0 & 1.6 & 0.386 & 3.208 & 0.036 \\ 
8.0 & 1.7 & 0.406 & 3.364 & 0.091 \\
\end{tabular}
\end{ruledtabular}
\end{table}

\begin{figure}[htb]
\includegraphics[width=\linewidth]{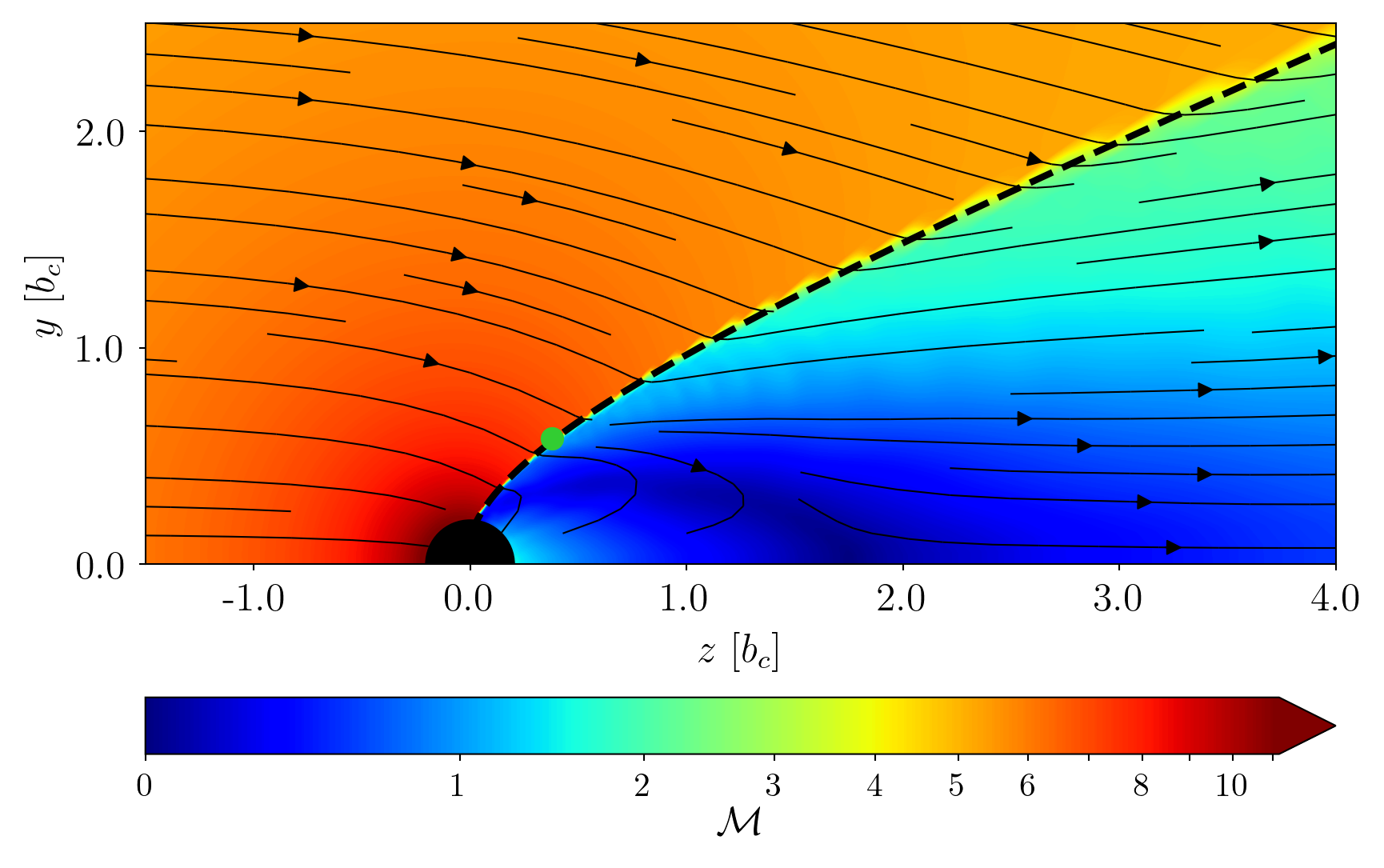}
\caption{Example of one of the numerical simulations
considered in this work for the particular wind parameters $\mathcal{M}_\infty=5$ and $\gamma=1.5$. The figure shows isocontours of the Mach number $\mathcal{M}$ and the resulting streamlines once a stationary state has been reached. The black, dashed line corresponds to the hyperbolic fit in 
Eq.~\eqref{eq:fit} 
(see table~\ref{tab:1}). 
Post-shock flow at the critical point (green dot) is approximately parallel to the $z$-axis.
}
\label{fig:S1}
\end{figure}

We have considered a set of 15 numerical simulations, performed using the non-relativistic version of {\tt aztekas}, with wind parameters in the range $1.3 \leq \gamma \leq 1.7$ and $2.5\leq \mathcal{M}_{\infty} \leq 8$ as listed in table~\ref{tab:1}. We adopted spherical coordinates and assumed axisymmetry. In all cases, the numerical domain was taken as $r\in[0.2\,b_c,\,5\,b_c]$, \mbox{$\theta\in[-\pi,\,\pi]$} and the simulations where left to run, starting from a uniform initial condition, up to a time \mbox{$T\simeq 60\,b_c/v_\infty $}, at which point a steady-state had been reached in all cases.  A regular grid of $256\times128$ points was used, together with a Courant number of 0.3. 
This choice of numerical parameters is based on the numerical analysis presented in \citep{2019MNRAS.487.3607T}. The obtained results are robust and reliable to within a factor of a few percent of a fully numerically converged result. Note that for the parameters considered here, the size of the event horizon is much smaller than the inner computational boundary.

In Fig.~\ref{fig:S1},
we show an example of the resulting steady-state accretion flow for $\mathcal{M}_\infty=5$ and $\gamma=1.5$, together with the corresponding hyperbolic approximation to the bow shock geometry as given in Eq.~\eqref{eq:fit}. 
Moreover, from the numerical results listed in table~\ref{tab:1}, we find the following second-order fitting formulae for the parameters $\beta_\infty$, $z_\infty$ and $z_0$ involved in this approximation:
\begin{align}
\beta_{\infty}  \simeq &\; -0.083 +1.151\;\! \gamma- 0.128 \;\! \mathcal{M}_{\infty}-0.318\;\! \gamma^2  \nonumber \\ 
    &\; -0.026 \;\!\gamma \;\! \mathcal{M}_{\infty} + 0.013\;\! \mathcal{M}_{\infty}^{\,2} , \\
z_{\infty} \simeq &\; +17.039 -23.364 \;\!\gamma -0.340\!\;\mathcal{M}_{\infty} +7.463\;\! \gamma^2\nonumber \\
    &\; +1.037\;\! \gamma \;\! \mathcal{M}_{\infty}  -0.105\;\! \mathcal{M}_{\infty}^{\,2} , \\
   z_0 \simeq &\; -1.001 +1.031\;\!\gamma -0.0381\;\! \mathcal{M}_{\infty} -0.179\;\!\gamma^2 \nonumber \\
   &\;- 0.006\;\!\gamma\;\!  \mathcal{M}_{\infty}  +0.004\;\! \mathcal{M}_{\infty}^{\,2} .
\end{align}

\begin{figure}[htb]
\includegraphics[width=\linewidth]{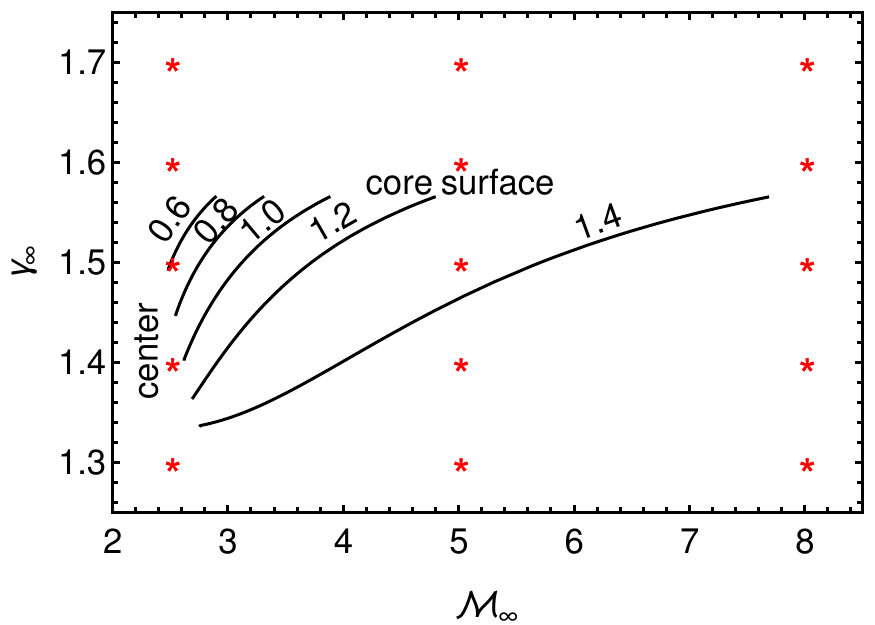}
\caption{Asymptotic adiabatic index versus asymptotic free fall Mach number for WDs with mass as given in the figure (in solar masses). The lines go from the core radius (top) to the center (left). Red stars indicate the parameters for which we performed wind simulations to determine the bow shock geometry (see table~\ref{tab:1}).}
\label{fig:S2}
\end{figure}

\section{Determining the induction length}\label{sec:ind}
In this section, we explain how we determined the induction length $\xi_{\perp}$ from initial conditions $\rho_1$, $v_{\!\perp 1}$ and $T_{\! 1}$. 
We proceed in three steps. First, we model the equation of state of hot shock-heated WD material. Second, we determine the post-shock state by solving the shock jump conditions. Third, we integrate the ZND equations. We assume uniformly initial temperature $T_{\!1}=10^7\,$K.

\subsection{Equation of state}\label{sec:eos}
\begin{figure}[htb]
\includegraphics[width=\linewidth]{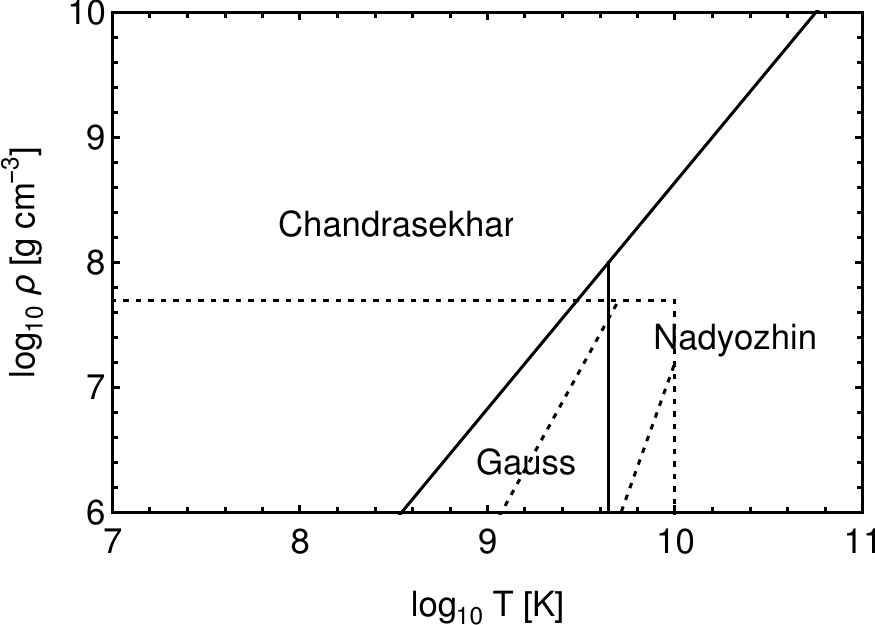}
\caption{Expansion formulae boundaries of highest accuracy for all thermodynamic quantities (including derivatives) as suggested by us (solid) and according to Ref.~\cite[fig. 12]{1996ApJS..106..171B} (dashed).}
\label{fig:S3}
\end{figure}

In shock-heated WD material, non-zero temperature effects in the equation of state $f(p,\rho,e)=0$ become important. We consider a gas made of electrons, positrons, ions and radiation. For the electron-positron component, we use the Nadyozhin equation of state \cite{1996ApJS..106..171B}, which is particularly well adapted for the technically difficult double intermediate regime where $ m_e\;\! c^2 \sim k_{\rm B}T$ and $\mu_{e}\sim k_{\rm B}T$, where $m_{e}$ is the electron mass an $\mu_{e}$ the electron chemical potential \cite{1999ApJS..125..277T}. For our parameter requirements, we use the Chandrasekhar expansion for the highly degenerate state, the Nadyozhin expansion for the ultra-relativistic state and the Gauss-Laguerre method for the intermediate state (see \cite{1996ApJS..106..171B} for a very detailed description). For the moment, we do not include the pair-plasma expansion. In all cases, we include ideal gas ions and radiation. We find a slightly different division in the $\rho-T$ parameter space more accurate than that suggested by \citet[fig. 12]{1996ApJS..106..171B}, see 
fig.~\ref{fig:S3}.

\subsection{Shock jump conditions}\label{sec:RH}
\begin{figure}[htb]
\includegraphics[width=\linewidth]{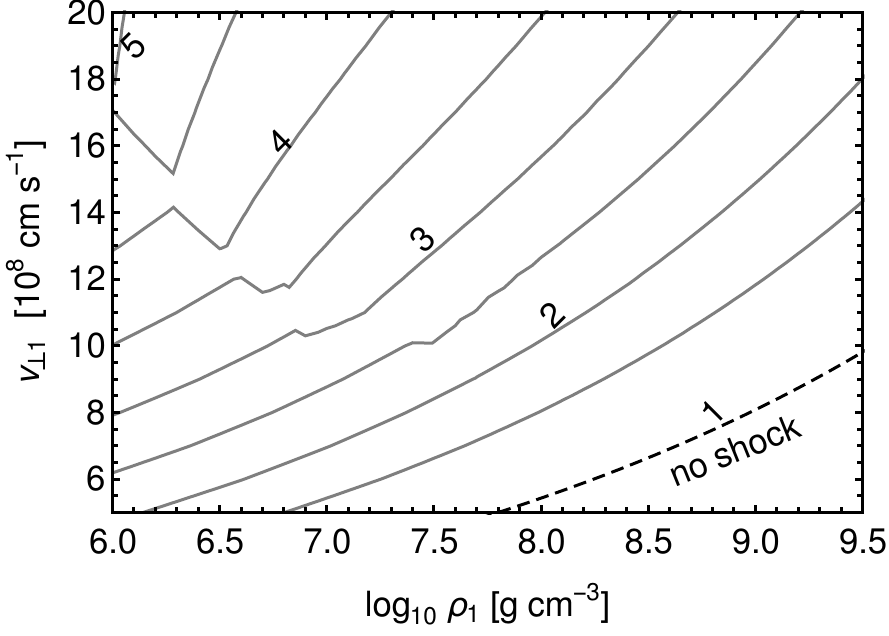}
\caption{Density ratio $\rho_2/\rho_1$ as function of pre-shock density and orthogonal velocity and assuming $T_{\!1}=10^7\,$K. The ``bump'' at low density and high velocity is due to electron-positron pair production and is expected.}
\label{fig:S4}
\end{figure}

\begin{figure}[htb]
\includegraphics[width=\linewidth]{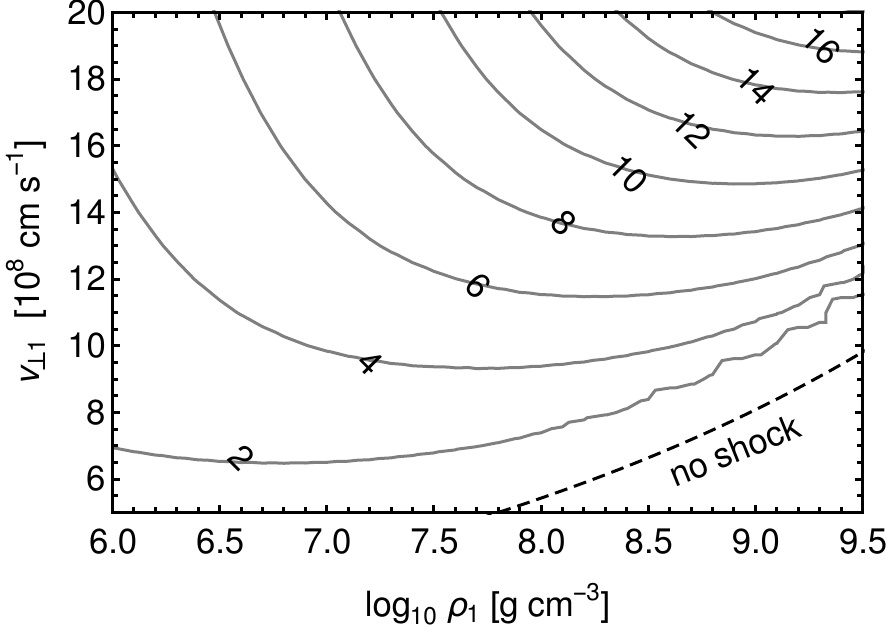}
\caption{Post-shock temperature $T_{\!2}/10^9$K as function of pre-shock density and orthogonal velocity and assuming $T_{\!1}=10^7\,$K.}
\label{fig:S5}
\end{figure}

The post-shock state is given by the Rankine-Hugoniot shock jump conditions
\begin{align}
    \rho_2\;\! v_{\perp 2} = &\; \rho_1\;\! v_{\perp 1}\,, \label{eq:RHv} \\
    v_{\parallel 2} =&\; v_{\parallel 1}\,, \\
 p_2 + \rho_2 \;\! v_{\perp 2}^{\,2} =&\; p_1 + \rho_1 \;\! v_{\perp 1}^{\,2}\,, \\
 e_2 + \frac{v_{\perp 2}^{\,2}}{2} =&\; e_1 + \frac{v_{\perp 1}^{\,2}}{2}\,, \label{eq:RHe}
\end{align}
where $v_{\perp}$ and $v_{\parallel}$ are the velocity components orthogonal and parallel to the shock, respectively. Eq.~(\ref{eq:RHv}) to Eq.~(\ref{eq:RHe}) can also be written in the more convenient form
\begin{align}
p_2 =&\; p_1 + \rho_1\;\! v_{\!\perp 1}^{\,2} \left(1-\frac{\rho_1}{\rho_2}\right), \label{eq:RH1} \\
e_2 =&\; e_1 + \frac{v_{\!\perp 1}^{\,2}}{2}\left(1-\frac{\rho_1^{\,2}}{\rho_2^{\,2}}\right). \label{eq:RH2}
\end{align}
Given an initial state ($\rho_1$, $v_{\perp 1}$, $T_{\!1}$), the post-shock state ($\rho_2$,$T_2$) is the solution of equations (\ref{eq:RH1}) and (\ref{eq:RH2}) together with the equation of state $f(p,\rho,e)=0$. We use a Newton-Raphson solver based on Jacobian inversion and making use of the Nadyozhin equation of state (see App.~\ref{sec:eos}). 
We require an accuracy of at least $10^{-4}$ in Eq.~(\ref{eq:RH1}) and Eq.~(\ref{eq:RH2}). The post-shock density and temperature are shown in 
figs.~\ref{fig:S4} and \ref{fig:S5},
respectively. In the parameter region where $v_{\perp 1} < a_1$ no shock occurs. The post-shock orthogonal Mach number is given by
\begin{equation}
    \mathcal{M}_{\perp 2} = \frac{v_{\perp 2}}{a_2} = \frac{\rho_1}{\rho_2}\;\! \frac{v_{\perp 1}}{a_2}
\end{equation}
where we used Eq.~(\ref{eq:RHv}) in the second equality, and $a_2$ is the post-shock sound speed determined by the equation of state.

\subsection{Induction length}\label{sec:ZND}
The reaction zone structure is governed by the Zeldovich-von Neuman-Doring (ZND) equations (see, for example, Refs.~ \citep{1989MNRAS.239..785K,1999ApJ...512..827G})
\begin{align}
    \frac{d\rho}{dt} =&\; \Big(\frac{\partial p}{\partial e}\Big)_{\!\!\rho}\;\!\frac{\dot{q}}{v_{\!\perp}^{\,2}\!-a^2} \,, \label{eq:ZND1}\\
    \frac{de}{dt} =&\; \frac{p}{\rho^2}\;\! \frac{d\rho}{dt} + \dot{q} \,,\label{eq:ZND2}\\
    \frac{d\xi_{\perp}}{dt} =&\; v_{\!\perp} = v_{\!\perp 2}\;\! \frac{\rho_2}{\rho}\,,\label{eq:ZND3}
\end{align}
where $v_{\!\perp 2}$ and $\rho_2$ are determined by the shock jump conditions (see App.~\ref{sec:RH}), $\dot{q}$ is the reaction rate, and $(\partial p/\partial e)_{\rho,\alpha_i}$ is the thermodynamic derivative at constant density and composition $\alpha_i$.

\begin{figure}[htb]
\includegraphics[width=\linewidth]{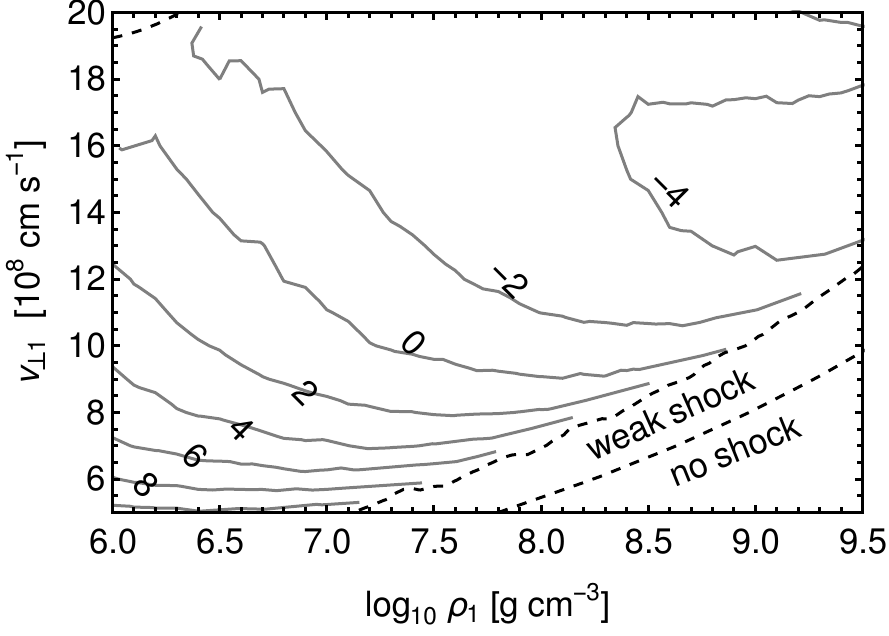}
\caption{Induction length, $\log_{10}(\xi_{\perp}/{\rm cm})$, as a function of pre-shock density $\rho_1$ and orthogonal velocity $v_{\!\perp 1}$, and assuming uniformly temperature $T_{\!1}=10^7\,$K.}
\label{fig:S6}
\end{figure}

We integrate the ZND Eqs. \eqref{eq:ZND1} to \eqref{eq:ZND3} using a 4th order Runge-Kutta method with adaptive time step. We find that convergence is attained as long as at most 10\% change is allowed in successive time steps. We stop the integration when the Chapman-Jouguet condition ($v=a$) is attained. As ingredients, we use the Nadyozhin equation of state code (see App.~\ref{sec:eos}) and the specific nuclear energy generation rate $\dot{q}$ of \cite{1988ADNDT..40..283C} including also electron screening (see, for example, Ref.~\cite{2019JCAP...08..031M} for details). The results are shown in 
fig.~\ref{fig:S6}
Parameter regions where no shock occurs ($v_{\!\perp 1}<a_1$) and parameter regions where only a weak shock occurs ($v_{\!\perp 2}>a_2$) are excluded. Our results agree with the carbon induction lengths calculated by \cite{1999ApJ...512..827G} for the parameters shown in their figs.~5 and 6.

When solving the RH equations and the ZND system, we assume throughout that the initial WD temperature is $10^7$K. The choice of this value is not critical, since the pressure in the degenerate state is independent of temperature \cite{2000ApJ...543..938T}.

\section{Fitting formula for the cross-section}\label{sec:fitting-cross}

We provide a fitting formula for the maximum radius for detonation ignition shown in 
Fig.~\ref{fig:cross}:
\begin{align}\label{eq:fitting-cross}
    \frac{R_m}{10^8\,{\rm cm}} \simeq  (2.46+0.48\;\! z) \Big[1-\Big(\frac{x-x_0}{1.44-x_0}\Big)^{\!2}\Big]^{1/2} 
\end{align}
where $x_0 \simeq 1.1600 - 0.0370\;\! z + 0.0100\;\!z^2-0.0026\;z^3$,  $x \equiv M_{\rm WD}/M_{\odot}$, and $z \equiv \log_{10}(M_{\rm BH}/10^{23}$g$)$. Formula \ref{eq:fitting-cross} is valid in the ranges $M_{\rm WD}\in[0.8,\,1.1]\,M_{\odot}$ and $\log_{10}(M_{\rm BH}/$g$)\in[21.5,\,25]$.

\end{document}